\documentclass[usegraphicx,usenatbib]{mn2e}

\newcommand{\xmm}{{\it XMM-Newton}}

\newcommand{\einstein}{{\it Einstein}}
\newcommand{\exosat}{{\it EXOSAT}}
\newcommand{\rosat}{{\it ROSAT}}
\newcommand{\xte}{{\it RXTE}}

\title[\xmm\ timing studies of DP Leo and WW Hor]
{First \xmm\ observations of strongly magnetic cataclysmic variables - II.
Timing studies of DP Leo and WW Hor}

\author[D. Pandel et al.]
{Dirk Pandel$^1$, France A. Cordova$^1$, Robert E. Shirey$^1$, Gavin Ramsay$^2$,\newauthor
 Mark Cropper$^2$, Keith O. Mason$^2$, Rudi Much$^3$, Dave Kilkenny$^4$\\
$^1$Department of Physics, University of California, Santa Barbara, CA 93106, USA\\
$^2$Mullard Space Science Lab, University College London, Holmbury St. Mary, Dorking, Surrey, RH5 6NT, UK\\
$^3$Astrophysics Division, ESTEC, 2200, AG Noordwijk, The Netherlands\\
$^4$South African Astronomical Observatory, PO Box 9, Observatory 7935, South Africa}

\pagerange{\pageref{firstpage}--\pageref{lastpage}}
\pubyear{2002}

\begin{document}

\maketitle

\label{firstpage}

\begin{abstract}

\xmm\ was used to observe two eclipsing, magnetic cataclysmic variables, DP Leo and WW Hor,
continuously for three orbital cycles each.
Both systems were in an intermediate state of accretion.
For WW Hor we also obtained optical light curves with the 
\xmm\ Optical Monitor and from ground-based observations.
Our analysis of the X-ray and optical light curves allows us to constrain
physical and geometrical parameters of the accretion regions
and derive orbital parameters and eclipse ephemerides of the systems.
For WW Hor we directly measure horizontal and vertical temperature variations
in the accretion column.
From comparisons with previous observations we find that
changes in the accretion spot longitude are correlated with the accretion rate.
For DP Leo the shape of the hard X-ray light curve is not as expected for optically thin emission,
showing the importance of optical depth effects in the post-shock region.
We find that the spin period of the white dwarf is slightly shorter than the orbital period
and that the orbital period is decreasing faster
than expected for energy loss by gravitational radiation alone.

\end{abstract}



\section{Introduction}

\begin{table*}
\caption[Observations summary]{
Details of the observations
(see also Section \ref{obsred})
}
\begin{tabular}{llcrrccccc}
\hline
Object & Instrument & \multicolumn{2}{c}{Beginning of} & Duration & Filter & Instrument & Timing & Aperture & Event \\
       &            & \multicolumn{2}{c}{Observation (UTC)} &     &        & mode   & resolution & radius   & pattern \\
\hline
DP Leo       & EPIC MOS 1      & 22 Nov 2000 & 4:55:02 & 22340 s  & Thin 1 & full window & 2.6 s   & 25'' & 0--12 \\
             & EPIC MOS 2      & '' & 4:54:54 & 22348 s  & Thin 1 & full window & 2.6 s   & 25'' & 0--12 \\
             & EPIC PN         & '' & 5:36:22 & 20034 s  & Thin 1 & full window & 0.073 s & 18'' & 0--12 \\
\hline
WW Hor       & EPIC MOS 1      & 4 Dec 2000 & 3:37:25 & 23543 s  & Thin 1 & full window & 2.6 s   & 25'' & 0--12 \\
             & EPIC MOS 2      & '' & 3:37:20 & 23548 s  & Thin 1 & full window & 2.6 s   & 25'' & 0--12 \\
             & EPIC PN         & '' & 4:18:48 & 21149 s  & Thin 1 & full window & 0.073 s & 18'' & 0--12 \\
             & Optical Monitor & '' & 3:30:09 & 9 $\times$ 2200 s & B & fast timing & 0.5 s & 3.2'' & --- \\
             & SAAO (1.0 m)    & 2 Dec 2000 & 19:41:35 & 7140 s   & R & --- & 60 s & --- & --- \\
\hline
\end{tabular}
\label{obstable}
\end{table*}

DP Leo and WW Hor are two eclipsing binary systems of type AM Her,
a class of magnetic cataclysmic variables also referred to as
polars because of their strongly polarized optical emission.
In these systems the strong magnetic field of the white dwarf primary
causes it to rotate synchronously with the orbital motion of the binary.
Due to the strong magnetic field, the accretion stream
from the Roche lobe filling secondary does not form a disk,
but rather follows the magnetic field lines onto the white dwarf's surface.
Slightly above the photosphere the accretion flow forms a shock
that heats the gas to temperatures above $\sim$10$^8$K.
The gas below the shock front then cools and settles onto the photosphere
\citep{2000SSRv...93..611W}.
Models of the accretion region suggest three major spectral components.
Thermal bremsstrahlung is emitted in the X-ray band
by the hot gas between the shock front and the surface.
The photosphere below the shock is heated to temperatures of $\sim$10$^5$K
by reprocessing of bremsstrahlung photons and by dense filaments in the accretion stream
\citep{1982A&A...114L...4K},
giving rise to a blackbody component that extends from the UV to the soft X-ray band. 
The strong magnetic field at the surface of the white dwarf causes the shock-heated electrons
to emit strongly polarized cyclotron radiation in the optical and IR bands. 
A comprehensive review of polars is given in
\citet{1990SSRv...54..195C}.

Much has been learned about the accretion region in polars from optical polarimetry.
These observations, however, are only able to reveal the physical conditions
that give rise to the cyclotron radiation.
To fully understand the structure of the post-shock region and to test dynamical models of the
accretion flow, the thermal bremsstrahlung and the blackbody component,
both visible in the X-ray band, need to be studied as well.
Past X-ray observations were limited by low sensitivity and a narrow X-ray bandpass.
With the high sensitivity of \xmm\ and its coverage of the 0.1--12 keV band
we can now obtain light curves and spectra with unprecedented quality.
This will allow us to independently study the bremsstrahlung and the blackbody component
and derive the physical and geometrical properties of the accretion regions.

In this paper we analyze X-ray and optical light curves of two polars 
in order to constrain the geometry of the accretion regions,
investigate properties of the X-ray emitting gas,
determine parameters of the binary systems,
and find changes in the accretion longitudes and orbital periods.
We present a detailed spectral analysis of these data in \citet{2001MNRAS.326L..27R}.

DP Leo, originally named E1114+182, was the first known eclipsing polar.
This binary system, which has an orbital period of 89.8 min,
was discovered serendipitously with the \einstein\ observatory
and quickly identified as a polar because of its
strongly modulated, polarized emission \citep{1985ApJ...293..303B}.
\citet{1994ApJ...437..436R} and \citet{1993MNRAS.261L..31B} later found
that the accretion spot longitude is changing by $\sim$2$^\circ$ per year,
suggesting a slightly asynchronous rotation of the white dwarf.
We have shown in \citet{2001MNRAS.326L..27R} that
the X-ray spectrum obtained with \xmm\ contains in addition
to the soft blackbody component a previously undetected hard bremsstrahlung component.
Using the X-ray spectrum we derived a white dwarf mass of $\sim$1.4$\ $M$_\odot$.

WW Hor (EXO 023432-5232), which has an orbital period of 115.5 min,
was discovered as a serendipitous X-ray source with \exosat\
and later identified as a polar \citep{1987A&A...175L...9B}.
Some of the system parameters were derived by \citet{1988MNRAS.234P..19B},
but orbital inclination and accretion spot latitude could not be determined.
A significant change of the accretion longitude over time was found by
\citet{1993MNRAS.261L..31B}.
As we have shown in \citet{2001MNRAS.326L..27R},
the \xmm\ spectrum does not contain the blackbody component
typically found in polars at soft X-ray energies.
From the X-ray spectrum we derived a white dwarf mass of $\sim$1.1$\ $M$_\odot$.


\section{Observations and analysis}

\subsection{Observations and data reduction}
\label{obsred}

\begin{figure*}
\includegraphics{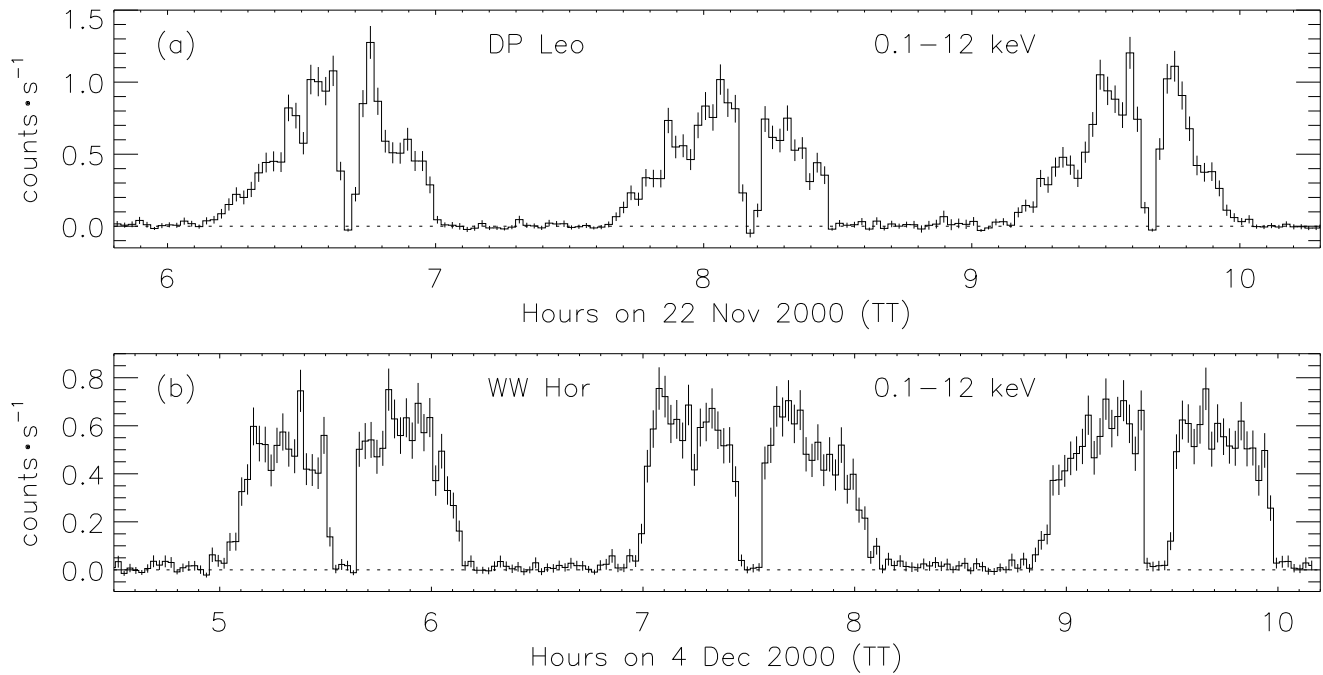}
\caption[DP Leo and WW Hor X-ray light curves]
{
X-ray light curves for DP Leo and WW Hor obtained with \xmm.
Figures show the combined EPIC MOS and PN count rates over three orbital cycles
in the 0.1--12 keV energy range.
The light curves can be clearly separated into
bright and faint phases each lasting about half of the orbit.
During the faint phase the accretion region is on the hemisphere
of the white dwarf facing away from us.
The narrow dip during the bright phase is the eclipse of the white dwarf
by the secondary star.
}
\label{lightcurves}
\end{figure*}

DP Leo and WW Hor were observed with \xmm\ for more than 20 ks each,
providing us with continuous X-ray data from all three EPIC instruments.
We did not use the data from the RGS instruments due to their low signal-to-noise ratio.
WW Hor was also observed with the \xmm\ Optical Monitor
and from the ground with the SAAO $1.0$-m telescope.
Details of the observations are shown in Table \ref{obstable}.

X-ray light curves were extracted using circular apertures
with the radii shown in Table \ref{obstable}.
For the background extraction we used regions near the source image and on the same CCD.
In the DP Leo data we found several periods with an increased background rate,
each lasting $\sim$1$\ $ks.
The largest increase occurred at the end of the observation 
so that we excluded the last 2 ks.
No significant fluctuations of the background rate were found in the WW Hor data.
After the initial data reduction we had for each target
continuous X-ray data covering three complete orbital cycles
available for our analysis.

During the WW Hor observation the \xmm\ Optical Monitor
was operating in fast timing mode with a time resolution of $0.5\ $s.
The Optical Monitor performed 9 consecutive B-band observations
with a duration of 2200 s each and separated by 300 s.
The location of the source image on the detector coincided with 
the edge of the central stray light ring \citep{2001A&A...365L..36M}.
This introduced uncertainties into the background calculation.


\subsection{Light curve analysis}

For our analysis we combined the EPIC MOS and PN data
by adding the identically binned and background subtracted
light curves from the three X-ray instruments.
We did not observe differences between the individual MOS and PN
light curves beyond those expected from statistical fluctuations.
Because of the still uncertain calibration of the encircled energy fraction 
we did not apply a correction for the aperture sizes
and the count rates quoted throughout the paper reflect only photons
detected inside these apertures.
The combined MOS and PN light curves
over the 0.1--12 keV band are shown in Fig. \ref{lightcurves}.
In order to improve the signal-to-noise ratio for our analysis and 
to reduce the fluctuations due to flaring
we folded the light curves on the known orbital periods
(5388 s for DP Leo \citep{1994ApJ...437..436R}
and 6929 s for WW Hor \citep{1990A&A...238..187B}).
The centre of the eclipse was chosen as orbital phase zero.
Folded X-ray light curves are shown in Fig. \ref{dpfold}a--c and \ref{wwfold}a--c.

The eclipse ingress and egress durations for both objects are very short,
near the limit of resolvability of the X-ray instruments
(Fig. \ref{dpeclipse} and \ref{wweclipse}).
We used a maximum likelihood method and the simple model
of a circular, flat, homogeneous, face-on emission region
to fit the eclipse profiles.
Ingress and egress durations quoted throughout the paper measure
the diameter of this disk-like region.
The width and centre of an eclipse were determined from the two points
where the flux is half of its average value immediately before or after the eclipse.
Photon arrival times were corrected to the solar system barycentre and the times 
of eclipse in this paper are shown as Barycentric Julian Dates (BJD) in 
Terrestrial Time (TT).
The eclipse centres in Tables \ref{dppars} and \ref{wwpars}
refer to the first eclipses in Fig. \ref{lightcurves}.
It should be noted that the time system used for \xmm\ data is TT 
and not UTC as stated in the documentation of the Science Analysis System
(XMM Helpdesk, priv. comm.).
Due to this incorrect information \citet{2002_SHS}, using the \xmm\ data for DP Leo,
overestimated the time of eclipse by $\sim$64$\ $s.



\subsection{DP Leo}
\label{obsdp}

\begin{figure*}
\includegraphics{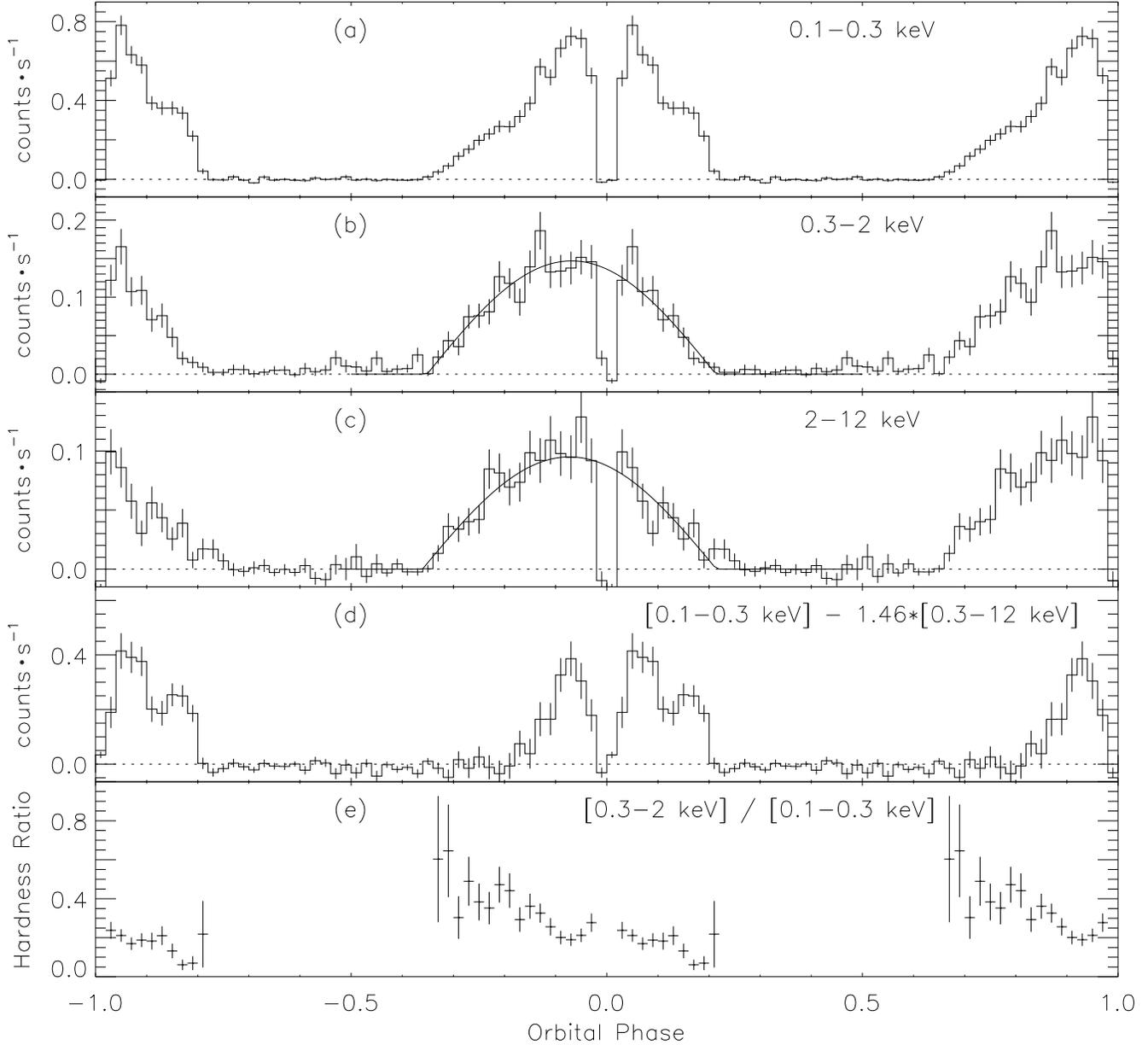}
\caption[DP Leo X-ray folded light curves]{
X-ray light curves for DP Leo folded on the orbital period of 5388 s (89.8 min).
{\it (a)} Soft blackbody component.
{\it (b,c)} Hard bremsstrahlung component;
Solid curve is a cosine fitted with variable width and centre.
{\it (d)} Residual of light curve (a) after the cosine profile shown in (b,c) has been subtracted
(see Section \ref{dpsoftem}).
{\it (e)} Hardness ratio; Points with large uncertainties have been omitted.
}
\label{dpfold}
\end{figure*}

As discussed in \citet{2001MNRAS.326L..27R},
the X-ray spectrum of DP Leo shows two distinct components,
a soft blackbody component dominating the flux below 0.3 keV
and a previously undetected hard bremsstrahlung component dominating above this energy.
The total flux observed with \xmm\ is dominated by the soft blackbody component
with more than 70 per cent of the X-rays in the energy range 0.1--0.3 keV.
Folded light curves for various energy ranges are shown in Fig. \ref{dpfold}.
By choosing 0.3 keV as the upper limit for the soft band
we almost completely separated
the soft blackbody and the hard bremsstrahlung component
with only a few per cent contamination remaining.

Surprisingly the light curves representing the bremsstrahlung component
(Fig. \ref{dpfold}b,c) do not show the top-hat shape that is expected for
an optically thin post-shock region.
Instead they appear to follow a cosine profile during the bright phase,
which is consistent with optically thick emission
(see Section \ref{dpbrems} for further discussion).
To the 0.3--12 keV light curve we fitted a cosine function
with variable width and peak position (solid curves in Fig. \ref{dpfold}b,c).
We obtained a good fit ($\chi_{red}^2=1.1$)
with a bright phase duration of $0.57(1)$ and a centre at phase $-0.070(3)$.
The latter value corresponds to an accretion spot longitude of $25^\circ\pm1^\circ$.
Here a longitude of $0^\circ$ is defined by the line between the two stars
and the positive sign indicates that the accretion spot is ahead of the secondary.

The light curve of the soft blackbody component (Fig. \ref{dpfold}a)
has a complex and asymmetric structure
not visible in the harder X-ray bands (Fig. \ref{dpfold}b,c).
The asymmetry can also be seen in the hardness ratio (Fig. \ref{dpfold}e)
which shows a softening of the spectrum throughout the bright phase.
Fig. \ref{lightcurves}a shows that some of this complex structure is due to 
short, random flares.
It appears that the flares are concentrated around the eclipse
and do not occur at the beginning of the bright phase.
We discuss the shape of the soft X-ray light curve further in
Section \ref{dpsoftem}.

\begin{figure}
\includegraphics{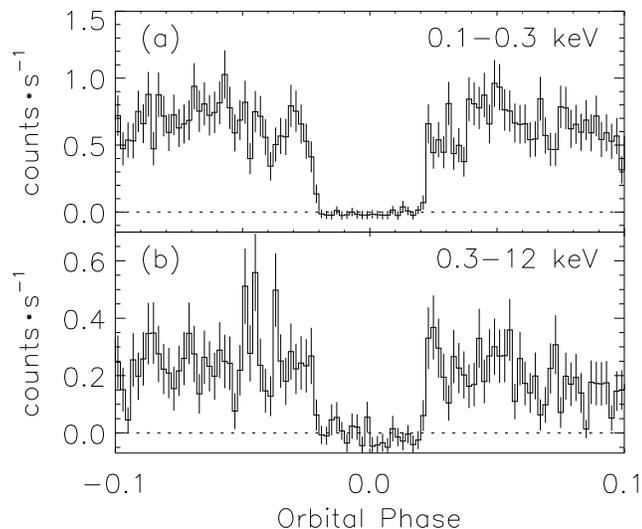}
\caption[DP Leo X-ray eclipse profiles]
{
X-ray light curves for DP Leo around the eclipse with
a bin size of 0.002 (10.8 s).
{\it (a)} Blackbody component.
{\it (b)} Bremsstrahlung component.
}
\label{dpeclipse}
\end{figure}

Folded X-ray light curves near the eclipse are shown in Fig. \ref{dpeclipse}.
The rapid decline to and rise from the eclipse is not resolved,
except for the ingress in the 0.1--0.3 keV band which lasted $40\pm10\ $s.
We find an upper limit of 11 s for the egress duration in the 0.1--0.3 keV band
and 40 s for the ingress and egress durations in the 0.3--12 keV band.
This is consistent with the result by \citet{1994ApJ...430..323S}
who measured an ingress/egress duration of $\sim$8$\ $s in the optical and UV.
A possible explanation for the difference between ingress and egress in the 0.1--0.3 keV band
is discussed in Section \ref{dpsoftem}.

\subsection{WW Hor}
\label{obsww}

\begin{figure*}
\includegraphics{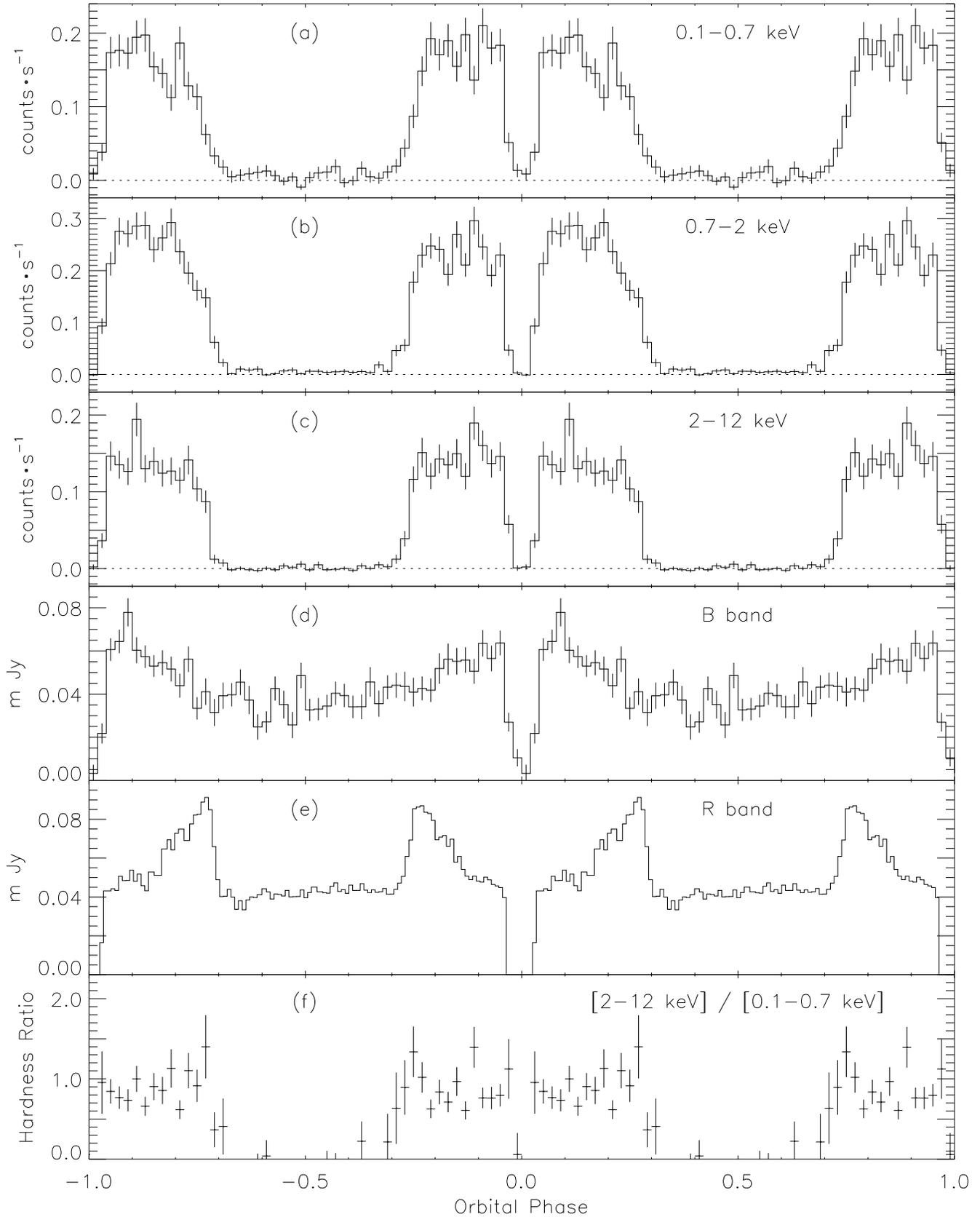}
\caption[WW Hor folded light curves]{
X-ray and optical light curves for WW Hor folded on the orbital period of 6929 s (115.5 min).
{\it (a--c)} Folded X-ray light curves obtained with \xmm.
{\it (d)} Folded B-band light curve obtained with the \xmm\ Optical Monitor.
{\it (e)} R-band light curve obtained with the SAAO 1.0-m telescope.
{\it (f)} Hardness ratio; Points with large uncertainties have been omitted.
}
\label{wwfold}
\end{figure*}

The folded X-ray light curves for WW Hor (Fig. \ref{wwfold}a--c)
roughly exhibit a top-hat shape.
This is expected for an optically thin emission region,
since the observed flux does not change with the viewing angle
unless part of the region becomes obscured by one of the stars.
In Fig. \ref{wwfold}a--c the transitions between bright and faint phases
are slower for the 0.1--0.7 keV band, indicating that the soft X-rays originate
from a larger region than the hard X-rays.
This can also be seen in the hardness ratio (Fig. \ref{wwfold}f)
which shows that the spectrum is softer near phase $\pm0.30$,
when the hard X-ray emitting inner region of the accretion spot is not visible,
and is harder near phase $\pm0.25$,
when only the soft X-ray emitting outer region is partially obscured.
The slower rise and decline in the soft X-ray light curve might also be due to
a non-negligible optical depth in the post-shock region.
It is generally thought that optical depth effects are not important for
bremsstrahlung, but \citet{2000NewAR..44...57C} argue that electron scattering
might produce a harder X-ray spectrum when the accretion column is viewed from
the side.
However, Fig. \ref{wwfold}f clearly shows a softer spectrum at the 
beginning and end of the bright phase.

In the light curves we determined the duration of the rise and decline near phase $\pm0.25$,
and the width and centre of the bright phase 
by fitting the profile expected for an optically thin, circular and
flat emission region.
Results are shown in Table \ref{wwpars}.
As discussed in Section \ref{wwaccgeo} these measurements can be used
to determine the size and height of the emitting regions.
The width of the bright phase was measured between the two points
where the flux is halfway between its high and its low level.
The centre of the bright phase at $0.002(2)$
corresponds to an accretion longitude of $-1^\circ\pm1^\circ$.

Fig. \ref{wwfold}d shows the folded B-band light curve
obtained with the \xmm\ Optical Monitor.
The higher flux level seen during the bright phase is most likely due to
thermal emission from the heated photosphere near the accretion region.
We measured a B magnitude of 20.2
during the faint phase and 19.6 at the peak.
The centre of the bright phase at 0.01(1)
coincides with that in the X-ray bands.

The R-band light curve in Fig. \ref{wwfold}e was measured
32 hours before the \xmm\ observation.
It shows the double-peaked profile typical for cyclotron radiation
from the accretion region.
Since cyclotron radiation is preferentially emitted perpendicular to the magnetic field lines,
the flux is highest near phase $\pm0.25$ when the accretion region is seen edge-on
and drops to nearly zero around the eclipse when the region is seen face-on.
The fast rise and decline near phase $\pm0.3$ is due to partial obscuration
of the accretion region by the white dwarf.
We determined the duration of this rise and decline and the width of the bright phase
using the same method as for the X-ray light curves.
Results are shown in Table \ref{wwpars}.
Since the companion star is not detected during the eclipse,
the constant flux level in Fig. \ref{wwfold}e seen during the faint phase
is mainly due to thermal emission from the white dwarf.
The contribution from the accretion stream or the heated side of the secondary is small
since no significant modulation is seen between phases 0.3 and 0.7.
We determined an R magnitude of 19.7 during the faint phase and 18.8 at the peaks.

\begin{figure}
\includegraphics{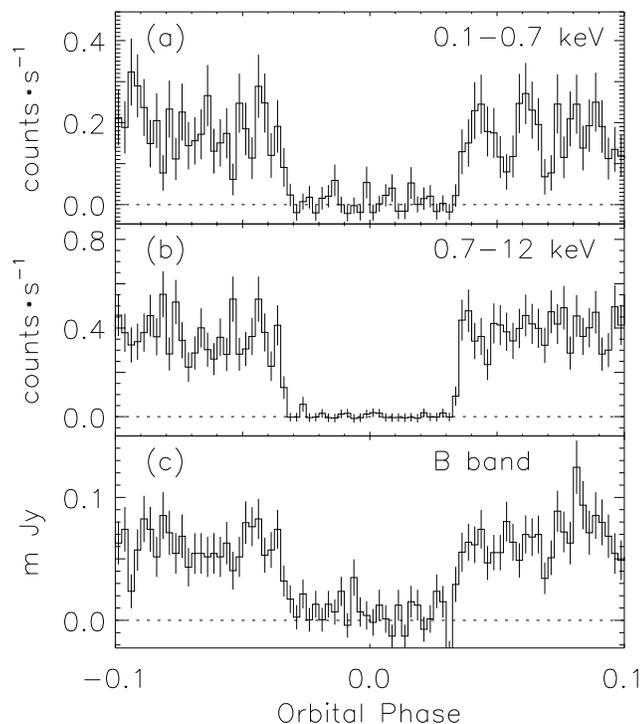}
\caption[WW Hor X-ray eclipse profiles]{
X-ray and optical light curves for WW Hor around the eclipse with
a bin size of 0.0025 (17.3 s).
}
\label{wweclipse}
\end{figure}

Fig. \ref{wweclipse} shows eclipse profiles in two X-ray bands and the B band.
For the ingress and egress durations we find in the 2--12 keV band an upper limit of 30 s
and in the 0.1--0.7 keV band a marginally non-zero value of $50\pm40\ $s.
The latter result is rather large compared to the upper limit of 7 s 
determined by \citet{1988MNRAS.234P..19B} from optical observations.
Our non-zero value may not be significant and
rather due to the strong fluctuations and low statistics.
In the B band we find non-zero ingress and egress durations of $\sim$40$\pm$20$\ $s.
This is expected since about half of the B-band emission
originates from the photosphere away from the accretion spot
and it roughly takes this long for the entire white dwarf to be eclipsed
\citep{1988MNRAS.234P..19B}.

\subsection{Oscillations}

The shock front in magnetic CVs may not be stable and
oscillate on a time scale of seconds.
In systems with a strong magnetic field, cyclotron cooling tends to suppress these
quasi-periodic oscillations (QPOs) \citep{1985ApJ...299L..87C}.
QPOs have been detected in the optical \citep{1989A&A...217..146L} but not as
yet in X-rays.
To search for oscillations we extracted X-ray light curves (0.1--12keV) with 0.01 s time bins
and performed Discrete Fourier Transforms.
We excluded the aforementioned time intervals of high background in the DP Leo observation
and did not perform a background subtraction.
No evidence for periodic or quasi-periodic oscillations
was found in either source.
For a strictly periodic signal we determined
an upper limit of $\sim$10\% for WW Hor and $\sim$22\% for DP Leo
over the frequency interval 0.1--10 Hz.
It is more difficult to give accurate upper limits for QPOs,
but we expect them to be comparable.
For V834 Cen \citet{2000PASP..112...18I} found with \xte\ data a similar
upper limit of $\sim$14\% at 90\%
confidence level for QPOs over the frequency range 0.2--1.2 Hz.
Our limits are not strong enough to test shock
oscillation models \citep{1999MNRAS.310..677S,2001A&A...373..211F}.           

\begin{table}
\caption{
Overview of the parameters we derived for DP Leo.
Values without explicit units are given in terms of orbital phase.
Numbers in parentheses following the values are the uncertainties of the last digits at a 1-sigma level.
}
\begin{tabular}{ll}
\hline
  Eclipse centre (BJD in TT)        & 2451870.77690(4) \\
  Eclipse period ($P_{ecl}$)        & 0.062362821(2) d \\
  Orbital conjunction (BJD in TT)   & 2451870.77682(4) \\
  Orbital period ($P_{orb}$)        & 0.062362820(2) d \\
  Period change ($\dot{P}_{orb}$)   & $-5.1(6)\cdot10^{-12}\ $s$\ $s$^{-1}$ \\
  Spin preriod ($P_{spin}$)         & 0.062362766(5) d \\
  Mass of secondary ($M_2$)       & $0.14(2)\ $M$_\odot$ \\
  Mass ratio ($q$)                & $0.09$--$0.16$ \\
  Orbital inclination ($i$)       & $81^\circ\pm1^\circ$ \\
  Accretion longitude             & $25^\circ\pm1^\circ$ \\
  Accretion longitude drift       & $2.3^\circ\pm0.2^\circ$ per year \\
  Bright phase duration (0.3--12 keV)  & $0.57(1)$ \\
  Bright phase centre (0.3--12 keV)    & $-0.070(3)$ \\
  Eclipse width (0.1-12 keV)           & 0.0437(6) \hfill ($235\pm3$ s) \\
\hline
\end{tabular}
\label{dppars}
\end{table}

\section{Discussion}


\subsection{DP Leo}


\subsubsection{Accretion state}

The X-ray peak flux we observed with \xmm\ corresponds
to a \rosat\ PSPC count rate of roughly 1 photon per second.
During a series of \rosat\ observations in 1992
DP Leo was found in a similar state of accretion
with a peak flux varying between 0.5 and 2 photons per second
\citep{1994ApJ...437..436R}.
Three months earlier an optical peak magnitude of R=17.9 was measured
\citep{1993MNRAS.261L..31B}.
From CCD images taken with the JKT 1.0-m telescope on La Palma (J. Etherton, priv. comm.)
one day after the \xmm\ observation
we obtain a similar magnitude of R$\sim$17.5-17.6 (at phase 0.13 near the suspected peak).
This is one magnitude fainter than the R=16.5 measured
during the high state in 1982 \citep{1985ApJ...293..303B}.
A lower state with a peak magnitude R$\sim$18.2 has also been observed
\citep{1994ApJ...430..323S}.
We conclude that during the \xmm\ observation
DP Leo was in an intermediate state of accretion.

\subsubsection{Orbital parameters}
\label{dpleoorbpar}

For cataclysmic variables the mass of the Roche-lobe filling secondary
can be inferred from the orbital period, given the mass-radius relationship of the star.
From mass-period relationships derived empirically
\citep{1995cvs..book.....W,1998MNRAS.301..767S}
and with evolutionary models \citep{2001ApJ...550..897H}
we estimate for DP Leo a secondary mass $M_2=0.14\pm0.02\ $M$_\odot$.
Here the assumption was made that DP Leo has not yet evolved past
the 80-min period minimum.
In \citet{2001MNRAS.326L..27R} we derived a white dwarf mass $M_1\sim1.4\ $M$_\odot$.
This is rather high and might be an overestimate.
With the conservative assumption $M_1\approx1.0$--$1.4\ $M$_\odot$
we obtain a mass ratio $q\approx0.09$--$0.16$.
For eclipsing CVs the orbital inclination can be derived from the duration of
the eclipse and the mass ratio \citep{1976ApJ...208..512C}.
With the eclipse duration of 235 s we obtain an inclination $i=81^\circ\pm1^\circ$.
This is consistent with the result $i=76^\circ\pm10^\circ$
by \citet{1985ApJ...293..303B}.
The accretion colatitude $\delta$ can in principle be derived from
the duration of the bright phase $\Delta\Phi$ using equation (\ref{equ1}) in Section \ref{wwaccgeo}.
However, since the accretion spot is most likely not small, the measured bright phase duration
is longer than the $\Delta\Phi_0$ needed in equation (\ref{equ1}).
It is possible to correct for this difference, but because of the uncertain accretion spot size
we were not able to obtain useful constraints for the colatitude.
The best available measurement from previous observations is
$\delta=103^\circ\pm5^\circ$ \citep{1985ApJ...293..303B}.

\begin{table}
\caption{
Overview of the parameters we derived for WW Hor.
Values without explicit units are given in terms of orbital phase.
Numbers in parentheses following the values are the uncertainties of the last digits at a 1-sigma level.
}
\begin{tabular}{ll}
\hline
  Eclipse centre (BJD in TT) & 2451882.73354(5) \\
  Eclipse period             & 0.0801990403(9) d \\
  Mass of secondary ($M_2$)  & $0.19(2)\ $M$_\odot$ \\
  Mass ratio ($q$)           & $0.14$--$0.22$ \\
  Orbital inclination ($i$)  & $84^\circ\pm2^\circ$ \\
  Accretion colatitude ($\delta$) & $64^\circ-104^\circ$ \\
  Accretion longitude         & $-1^\circ\pm1^\circ$ \\
{\it Optical magnitudes} \\
  \ \ \ B band (peak/faint phase) & 19.6 / 20.2 \\
  \ \ \ R band (peak/faint phase) & 18.8 / 19.7 \\
{\it Bright phase width} \\
  \ \ \ X-rays (0.1--0.7 keV)    & $0.515(7)$ \\
  \ \ \ X-rays (2--12 keV)       & $0.530(4)$ \\
  \ \ \ R band                   & $0.551(4)$ \\
{\it Bright phase centre} \\
  \ \ \ X-rays (0.1--12 keV)       & $0.002(2)$ \\
  \ \ \ B band                     & $0.01(1)$ \\
{\it Bright phase rise/decline width} \\
  \ \ \ X-rays (0.1--0.7 keV) & 0.09(2) / 0.12(2) \\
  \ \ \ X-rays (2--12 keV)    & 0.06(1) / 0.06(1) \\
  \ \ \ R band                & 0.04(1) / 0.03(1) \\
{\it Width of eclipse} \\
  \ \ \ X-rays (0.1--12 keV)  & $0.0682(7)$ \hfill ($473\pm5$ s) \\
  \ \ \ B band                & $0.068(2)$ \hfill ($470\pm14$ s) \\
\hline
\end{tabular}
\label{wwpars}
\end{table}


\subsubsection{Eclipse ephemeris and orbital period change}

In Fig. \ref{dpephem}a we compare our measurement of the time at which the 
eclipse occurred with previous results by
\citet{1994ApJ...437..436R,1994ApJ...430..323S,1988prco.book...85S,1985ApJ...293..303B}.
To account for leap seconds accumulated over the past 20 years we converted
the eclipse timings from these papers, originally in HJD (UTC), to BJD (TT).
Fig. \ref{dpephem}a shows the deviation of the measured time of eclipse from a 
linear ephemeris (dashed line) that was obtained by fitting only the data points 
between 1979 and 1993.
During the \xmm\ observation in November 2000 the eclipse occurred 
$\sim$70$\ $s earlier than predicted by the linear ephemeris.
This is clear evidence that the orbital period $P_{orb}$ of the system has been 
decreasing over the past 20 years.
Marginal evidence for a changing $P_{orb}$ was found by 
\citet{1994ApJ...430..323S}.

The solid curve in Fig. \ref{dpephem}a shows the result
of a quadratic fit to all data points including our measurement.
From this fit we obtain an improved
eclipse ephemeris that now contains a quadratic term
to account for the non-zero $\dot{P}_{orb}$:
\begin{eqnarray*}
\mbox{BJD (TT) }\ 2451870.77690(4) & + & 0.062362821(2) \cdot E  \\
& - & 1.6(2)\cdot10^{-13} \cdot E^2
\end{eqnarray*}
The quadratic term corresponds to an orbital period change
$\dot{P}_{orb}=-5.1(6)\cdot10^{-12}\ $s$\ $s$^{-1}$,
which gives a characteristic time scale
$P_{orb}/\dot{P}_{orb}=-3.3(4)\cdot10^7\ $yrs.
This is one order of magnitude shorter than the time scale of $-5\cdot10^8\ $yrs
expected for angular momentum loss by gravitational radiation alone
\citep{1994Ap&SS.211...61W}.

The eclipse ephemeris slightly differs from the ephemeris for orbital conjunction
because the accretion region is $25^\circ$ ahead of the secondary.
Using the result by \citet{1994ApJ...430..323S} that it takes 51 s to eclipse the entire white dwarf
and with the orbital parameters in Table \ref{dppars},
we estimate that during the \xmm\ observation orbital conjunction occurred 8 s 
before the centre of the eclipse was seen.
By applying similar corrections to all previous measurments
we obtain an ephemeris for orbital conjunction:
\begin{eqnarray*}
\mbox{BJD (TT) }\ 2451870.77682(4) & + & 0.062362820(2) \cdot E  \\
& - & 1.6(2)\cdot10^{-13} \cdot E^2
\end{eqnarray*}


\subsubsection{Accretion longitude changes}
\label{dpacclong}

\begin{figure}
\includegraphics{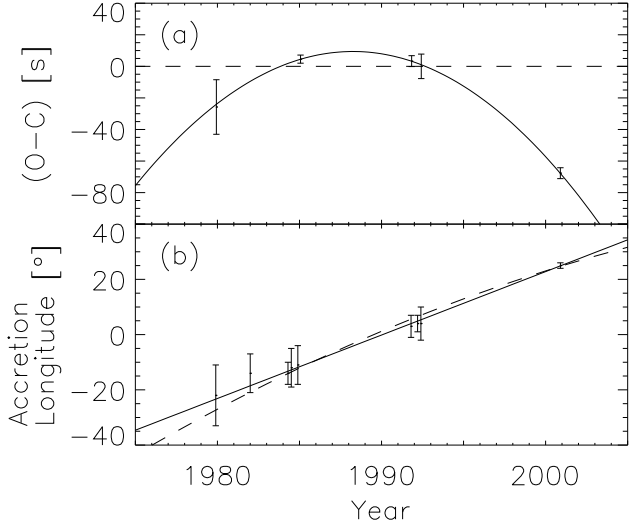}
\caption[DP Leo period and accretion longitude changes]{
DP Leo:
{\it (a)} Deviation of the time when the eclipse occurred from a linear ephemeris (dashed line)
that was obtained by fitting only the data points between 1979 and 1993.
The solid curve is the best fit of a quadratic ephemeris to all data points.
{\it (b)}$\ $Accretion longitudes measured over the past 20 years.
The solid line is a linear fit with a slope of $2.3^\circ\pm0.2^\circ$ per year.
The dashed curve shows a fit corrected for the change in orbital period.
}
\label{dpephem}
\end{figure}

The simple and symmetric shape of the hard X-ray light curves (Fig. \ref{dpfold}b,c)
allowed us to determine the accretion longitude with high accuracy ($25^\circ\pm1^\circ$).
In Fig. \ref{dpephem}b we compare our result with earlier measurements by
\citet{1994ApJ...437..436R,1994ApJ...430..323S,1993MNRAS.261L..31B,1985ApJ...293..303B}.
The data suggest that the accretion longitude has been increasing linearly with time
over the past 20 years.
From a linear fit we find a longitude drift of $2.3^\circ\pm0.2^\circ$ per year
(solid line in Fig. \ref{dpephem}b).
This drift is likely caused by a slightly asynchronous rotation of the white dwarf.
If the accretion spot is fixed relative to the magnetic poles,
then the spin period of the white dwarf must be $5.9(5)\ $ms shorter than the orbital period.
This corresponds to a relative deviation of $1.1(1)\times10^{-6}\ $s$\ $s$^{-1}$
and a synodic period of $\sim$160 years.
The actual deviation between the two periods might be somewhat different,
since changes in the orientation of the magnetic axis are
likely to affect the location of the accretion spot.

Magnetic interactions between the two stars are expected to
spin down the white dwarf toward synchronism on a time-scale of $\sim$200 years
\citep{1999A&A...343..132C}.
However, for DP Leo the orbital period is decreasing faster than the spin period is increasing
so that synchronism will be achieved on a shorter time-scale.
The dashed curve in Fig. \ref{dpephem}b shows a fit to the accretion longitudes
taking into account the change in orbital period.
Here we assumed that the spin period $P_{spin}$ and
the orbital period change $\dot{P}_{orb}$ are constant.
We find that in December 2000 the spin period
was $8.6\times10^{-7}\ $s$\ $s$^{-1}$ shorter than the orbital period
and that DP Leo will achieve synchronism in the year 2030.
However, the white dwarf spin will not remain sychronized
as the orbital period continues to decrease,
leading to an under-synchronous rotation
and a decreasing spot longitude.


\subsubsection{Hard X-ray light curves}
\label{dpbrems}

The X-ray light curves of the bremsstrahlung component (Fig. \ref{dpfold}b,c)
closely follow a cosine profile during the bright phase.
This is an unexpected result, since the region emitting the bremsstrahlung
is thought to be optically thin and the light curves should have a top-hat shape.
A similar cosine profile was seen before with \rosat\ and \einstein\
\citep{1994ApJ...437..436R}.
However, these observatories most likely did not detect the bremsstrahlung component
but rather the dominating soft blackbody component:
this emission originates from the optically thick photosphere
and is expected to produce a cosine-shaped light curve.

\citet{1983ApJ...268..291I} calculated X-ray light curves for
the bremsstrahlung component in polars.
They showed that for a system with orbital inclination $i\approx90^\circ$
the light curve has a top-hat shape if the accretion region is near the equator,
but can also have a shape closer to a cosine if accretion occurs near the rotational poles. 
In the latter case the post-shock region is always viewed from the side and,
since it is much more horizontally than vertically extended,
might always appear optically thick.
However, the accretion rate in DP Leo was $\sim$10$^2$ times lower than the 
rates assumed by the authors, so that the optical depth might be significantly 
less.

Unfortunately we are unable to constrain the colatitude of the
accretion region (Section \ref{dpleoorbpar}).
If accretion does occur close to the lower rotational pole 
($\delta\approx180^\circ$), then the model by \citet{1983ApJ...268..291I}
might be able to explain the observed cosine profile.
However, it is not clear why the accretion region would have migrated this far
from its 1984 location at $\delta\approx100^\circ$ \citep{1985ApJ...293..303B}.
The considerable shift in accretion longitude by $\sim$40$^\circ$ since 
1984 suggests that significant changes in the entire accretion geometry 
have occurred.
However, a shift in colatitude would have caused a change in the bright phase 
duration which is not observed.
If the accretion spot is still near the equator as in 1984,
then optical depth effects as discussed in \citet{2000NewAR..44...57C}
are also important when the accretion region is seen face-on.

The observed hard X-ray light curve cannot be easily explained by an
optically thin, but horizontally extended emission region.
Since the flux is at its highest level for only a short period
($\le$0.2 in phase),
an optically thin region would have to be partially obscured during 
most of the bright phase.
For this to be possible an accretion region near the equator would have to be 
almost as large as the white dwarf.
The accretion region in DP Leo is much smaller than this, only $\sim$15\% of 
the white dwarf diameter \citep{1993MNRAS.261L..31B}.
An optically thin region of this size can only produce the observed light curve 
if it is located at high colatitudes ($\delta\approx170^\circ$).
As discussed in the previous paragraph, the accretion region in DP Leo 
is more likely close to $\delta\approx100^\circ$.

Optical depth effects in the accretion column are likely to affect the
X-ray spectrum \citep[e.g.][]{1998MNRAS.297..526H}.
However, current models do not consider optically thick post-shock regions and
it is difficult to predict how the spectrum will deviate from the optically thin
case.
One might expect to see a blackbody-like spectrum that is modified somewhat due
to the inhomogeneous temperature and density distribution and that changes
with the viewing angle.
Yet for DP Leo the X-ray spectrum above 0.3 keV is well fit by an optically thin
bremsstrahlung model \citep{2001MNRAS.326L..27R} and the spectral slope
apparently does not vary during the bright phase (Fig.$\ $\ref{dpfold}b,c).
It is generally thought that free-free absorption is negligible
for X-rays in the post-shock region.
However, the optical depth due to electron scattering might be
significant \citep{2000NewAR..44...57C,1992MNRAS.254..493R}.
In a post-shock region with an optical depth for electron scattering larger
than unity and with negligible free-free absorption, the photons produced by
bremsstrahlung are typically not reabsorbed, but are scattered until they
escape from the region.
Electron scattering is wavelength independent and, if the photon energies
are not significantly changed by the thermal motion of the electrons,
the energy distribution of the escaping photons should be very similar
to their original bremsstrahlung spectrum.
Because the region is optically thick for electron scattering, the observed
flux depends mainly on the projected area and should vary like a cosine,
just as it does for DP Leo.
So if electron scattering is the major source of opacity,
it is possible to obtain the spectrum of optically thin bremsstrahlung
even if the post-shock region is optically thick.
Of course, for a more realistic picture other effects like cyclotron
cooling, inverse compton scattering and the accretion column structure need to
be considered.


\subsubsection{Soft X-ray light curve}
\label{dpsoftem}

The soft X-ray light curve (Fig. \ref{dpfold}a) has a complex and asymmetric 
shape very different from the simple cosine profile seen in the hard X-ray bands 
(Fig. \ref{dpfold}b,c).
Strong asymmetries in the X-ray and optical light curves of DP Leo have been 
observed before \citep{1994ApJ...437..436R,1993MNRAS.261L..31B}.
The hardness ratio in Fig. \ref{dpfold}e shows a softening of the spectrum
throughout the bright phase.
These variations might be due to photoelectric absorption of the soft X-rays
by an asymmetric accretion curtain.
A column density $N_H\sim 10^{20}{\rm cm}^{-2}$ is required
to explain the amplitude of the observed changes.

As an alternative explanation we suggest that the structure in the soft X-ray 
light curve might be caused by additional accretion regions 
only visible at energies below $0.3\ $keV.
Almost all of the emission in the 0.1--0.3 keV band is blackbody radiation 
from the photosphere below the shock \citep{2001MNRAS.326L..27R}.
Since the main accretion region is small
\citep[$\sim$0.15$\ R_{WD}$; e.g.][]{1993MNRAS.261L..31B}
and near the equator, the soft X-ray light curve in Fig. \ref{dpfold}a 
should be cosine-shaped similar to the light curves in Fig. \ref{dpfold}b,c.
Such a cosine profile for the soft X-ray emission was seen before with \rosat\ 
and \einstein\ \citep{1994ApJ...437..436R}.
Fig.$\ $\ref{lightcurves}a shows that some of the deviation from the 
expected cosine profile is due to short flares.
Yet the asymmetry seen in Fig.$\ $\ref{dpfold}a is present for all three orbital 
cycles and it appears that the flares are concentrated around the eclipse.
If only one accretion region is present, we would expect to see a symmetric 
light curve and the flares should be more evenly spread over the bright phase.
However, the soft X-ray light curve could be explained with a second
accretion region that is at higher colatitudes than the main region and has a 
shorter bright phase.
Unfortunately the \xmm\ observation covered only three orbital cylces and it is 
difficult to know how permanent the features in Fig. \ref{dpfold}a are.

If the structure in the soft X-ray light curve is indeed caused by additional 
accretion regions, it is possible to derive their location on the white dwarf.
The soft X-ray emission from the main accretion region is expected to 
produce a cosine-shaped light curve with the same bright phase duration of 
0.57 that was found in the 0.3--12 keV band.
We therefore subracted this cosine profile from the soft X-ray light curve
(Fig. \ref{dpfold}a), which would ideally remove the contribution from
the main region.
The normalization was chosen such that the residual is as small 
as possible but not negative beyond statistical fluctuations.
The residual light curve (Fig. \ref{dpfold}d) shows a broad peak centred around 
the eclipse and a narrower peak centred at phase 0.17.
These peaks could be interpreted as emission from two regions near 
the lower rotational pole.
By measuring the peak width and centre we can determine the locations 
of the associated regions with equation (\ref{equ1}) in Section \ref{wwaccgeo}.
We find for the larger peak a longitude of $-2^\circ$ and a colatitude of 
$>164^\circ$ and for the smaller peak a longitude of $-61^\circ$ and a 
colatitude of $\sim$172$^\circ$.
In comparison the main accretion region was found at $25^\circ$ longitude and 
the colatitude was not constrained (Sections \ref{obsdp} and \ref{dpleoorbpar}).

If these additional emission regions exist, it should be possible to 
distinguish them in the eclipse profile from the main accretion region.
With the orbital parameters in Table \ref{dppars} we find
that during ingress the limb of the secondary moves across the white dwarf
at an angle of about $+45^\circ$ with respect to the rotational axis
while during egress it does so at an angle of $-45^\circ$.
With a colatitude $\delta=103^\circ$ for the main region 
\citep{1985ApJ...293..303B} and using the result by \citet{1994ApJ...430..323S}
that it takes 51 s for the secondary to eclipse the entire white dwarf,
we estimate that during ingress the second region would be eclipsed 20 s
earlier than the main region,
but during egress it would be eclipsed 4 s later.
This is in agreement with the eclipse profile in Fig.$\ $\ref{dpeclipse}a,
which shows a significantly longer ingress than egress duration.
In the figure a time difference of 20 s corresponds to roughly 2 bins.
However, we cannot exclude the possibility that the longer ingress duration
was caused by statistical fluctuations or flaring.
The model of a second emission region near the rotational pole
and in longitude behind the main region
could also account for the asymmetries seen in the X-ray and optical
light curves from previous observations
\citep{1994ApJ...437..436R,1993MNRAS.261L..31B}.


\subsection{WW Hor}


\subsubsection{Accretion state}
\label{wwaccstate}

During the \xmm\ observation of WW Hor
we measured optical peak magnitudes of $B=19.6$ and $R=18.8$.
Similar values of $B=19.6$ and $R=18.3$
were found in a 1992 observation \citep{1993MNRAS.261L..31B}.
One week earlier the \rosat\ PSPC measured a peak flux of
$0.08$ photons per second \citep{1994MNRAS.271..733T}.
This is comparable to the \xmm\ flux which corresponds
to a \rosat\ PSPC count rate of $\sim$0.06 photons per second.
WW Hor was apparently in similar accretion states 
during the \xmm\ and the 1992 observation.
In 1987 WW Hor was found in a higher state
with magnitudes of $B=18.6$ and $R=17.1$ \citep{1988MNRAS.234P..19B}.
Accretion states lower than in December 2000 have not been observed,
but since strong cyclotron emission is visible in our R-band data (Fig. \ref{wwfold}e)
we conclude that WW Hor was in an intermediate state of accretion.


\subsubsection{Orbital parameters}
\label{wworbpar}

The orbital parameters of WW Hor can be derived in a similar way
as shown in Section \ref{dpleoorbpar}.
From the 115.5-min orbital period we find a secondary mass
$M_2=0.19\pm0.02\ $M$_\odot$ and with the white dwarf mass $M_1\approx0.9$--$1.3\ $M$_\odot$
\citep{2001MNRAS.326L..27R} a mass ratio $q\approx0.14$--$0.22$.
By combining this with the eclipse duration of 473 s
we obtain an orbital inclination $i=84^\circ\pm2^\circ$.
The accretion colatitude $\delta$ can be derived from
the bright phase duration $\Delta\Phi$ using equation (\ref{equ1}) in Section \ref{wwaccgeo}.
By measuring $\Delta\Phi$ between the points where
the flux is halfway between its high and its low level,
the complications from the uncertain accretion spot size
we encountered for DP Leo can be avoided.
As shown in Table \ref{wwpars}, the bright phase widths differ between
energy bands, which is probably due to different heights above the surface.
If we assume that $\Delta\Phi_0=0.515$, i.e. the 0.1--0.7 keV flux originates
from a region at zero height, we find an accretion spot colatitude
$\delta\approx66^\circ\pm13^\circ$.
Since this assumption might not be valid,
we use the R-band light curve (Fig. \ref{wwfold}e)
to obtain a more reliable constraint for $\delta$.
The flux from cyclotron emission around the eclipse is less than $\sim$10\%
of its peak value.
From the $\sin^2\alpha$ dependence of the flux we therefore know
that around the eclipse the angle $\alpha$ between the
line of sight and the magnetic field lines is less than $20^\circ$.
If the magnetic field is perpendicular to the surface,
this requires that $\delta$ and $i$ differ by at most $20^\circ$,
i.e. $\delta\approx64^\circ-104^\circ$.


\subsubsection{Eclipse ephemeris}

The eclipse observed with \xmm\ occurred as predicted by the latest ephemeris
\citep{1990A&A...238..187B} which had an accumulated uncertainty of $\sim$40$\ $s.
We obtain an updated ephemeris
by converting the mid-eclipse times in table 2 of \citet{1990A&A...238..187B}
from HJD (UTC) to BJD (TT) and combining them with our measurement:
\[ \mbox{BJD (TT) }\ 2451882.73353(5)+0.0801990403(9)\cdot E \]
Since no change in orbital period is seen, we can place an upper limit of
$5\cdot10^{-12}\ $s$\ $s$^{-1}$ on $\dot{P}_{orb}$ or a lower limit of
$4\cdot10^7\ $yrs on $P_{orb}/\dot{P}_{orb}$.


\subsubsection{Accretion longitude changes}

From the X-ray light curves we derived an accretion longitude of $-1^\circ\pm1^\circ$.
\citet{1993MNRAS.261L..31B} determined a similar value of $-3.6^\circ\pm2.5^\circ$ in 1992,
when WW Hor was in a similar intermediate state of accretion.
The authors found a significantly different value of $11.2^\circ\pm1.8^\circ$ for 1987,
when WW Hor was in a high accretion state and brighter by 1.7 magnitudes in R.
These data are not consistent with a linear time dependence of
the longitude, but rather suggest a correlation with the accretion rate.
The observed longitude variations might be due to a changing location of the
accretion spot on the white dwarf's surface,
while the orientation of the magnetic axis relative to the secondary remains fixed.


\subsubsection{Accretion region geometry}
\label{wwaccgeo}

\begin{table}
\caption
{
Horizontal and vertical extent of the emission region in WW Hor
(see Section \ref{wwaccgeo}).
{\it 2$^{nd}$ column} is the horizontal extent in a longitudinal direction.
{\it 3$^{rd}$}--{\it 5$^{th}$ column} is the height in units of white dwarf
radius for three assumed colatitudes $\delta$ inside the allowed range
$64^\circ-104^\circ$ from Section \ref{wworbpar}.
}
\begin{tabular}{lcccc}
\hline
Energy band & Extent in & \multicolumn{3}{c}{Height in units of $R_{WD}$} \\
            & longitude & $\delta=64^\circ$ & $\delta=84^\circ$ & $\delta=104^\circ$ \\
\hline
0.1-0.7 keV & $38^\circ\pm6^\circ$ & 0.000 & 0.001 & 0.002 \\
2-12 keV    & $22^\circ\pm4^\circ$ & 0.001 & 0.003 & 0.007 \\
R band      & $13^\circ\pm4^\circ$ & 0.005 & 0.011 & 0.016 \\
\hline
\end{tabular}
\label{wwgeo}
\end{table}

As discussed in \citet{2001MNRAS.326L..27R},
the X-ray spectrum of WW Hor does not appear to
contain the soft blackbody component usually seen in polars.
The light curves shown in Fig. \ref{wwfold}a--c therefore
represent different energy ranges of the bremsstrahlung component.
Their top-hat like shape is as expected for the optically thin
bremsstrahlung emission and in agreement with models of the post-shock region
\citep{2000NewAR..44...57C,1983ApJ...268..291I}.
By measuring the shape of the X-ray and R-band light curves
we are able to constrain the horizontal and vertical extent of the emission region.

In the idealized case of an optically thin and point-like emission region at zero height
the light curve is expected to have a perfect top-hat shape
and the duration of the bright phase $\Delta\Phi_0$ is given by
\begin{equation}
\label{equ1}
\cot i \cdot \cot \delta = - \cos ( 180^\circ \cdot \Delta\Phi_0 )
\end{equation}
Here $i$ is the orbital inclination, $\delta$ is the colatitude of the emission region
and $\Delta\Phi_0$ has units of orbital phase.
If the emitting region is not point-like but extended horizontally,
the transition between bright and faint phases is more gradual
than for a top-hat profile.
The duration of this transition is equal to the time it takes the accretion 
region to move over the horizon of the white dwarf
and therefore is a direct measure for the angular size of the region in 
a longitudinal direction.
The angular size does not depend on the width of the bright phase
which is a measure for the height of the region (see next paragraph).
Table \ref{wwgeo} ($2^{\rm nd}$ column) shows the angular sizes we obtain by 
converting the rise and decline widths from Table \ref{wwpars} into degrees.
We find that the soft X-ray emitting region with a size of $\sim$40$^\circ$
is about twice as large as the region in which the hard X-rays are produced.
This suggests that the temperature in the post-shock region
is not uniform across the spot but rather increases toward the centre.
The emission in the R-band originates from an even smaller region,
indicating that efficient production of cyclotron radiation requires
even higher temperatures than the emission of hard X-rays via bremsstrahlung.

For an extended emission region equation (\ref{equ1}) is still valid
if the width of the bright phase $\Delta\Phi_0$ is measured between the two points
where the flux is halfway between its high and its low level.
However, for a region with a significant height above the surface the measured width $\Delta\Phi$
is longer than the $\Delta\Phi_0$ predicted by equation (\ref{equ1}).
Given the inclination $i$ and colatitude $\delta$ 
we can calculate $\Delta\Phi_0$ with equation (\ref{equ1})
and then estimate the height $H$ from the measured width
$\Delta\Phi$ using the relation
\begin{equation}
\label{equ2}
H = \frac{1}{2}\ \pi^2\cdot( \Delta\Phi - \Delta\Phi_0 )^2\cdot\sin^2i\cdot\sin^2\delta\cdot R_{WD}
\end{equation}
which is valid for $H\ll R_{WD}$.
Here $\Delta\Phi$ and $\Delta\Phi_0$ have units of orbital phase
and $R_{WD}$ is the radius of the white dwarf.
Table \ref{wwgeo} shows the values of $H$ we obtain from the bright phase widths
in Table \ref{wwpars} for three assumed colatitudes $\delta$ inside the allowed range
from Section \ref{wworbpar}.
The cyclotron radiation originates from a height of $\sim$0.01$\ R_{WD}$
while the bremsstrahlung component is produced much closer to the surface.
We also find that the soft X-rays are emitted at a lower height than the hard X-rays.
This result is consistent with the picture that the gas in the post-shock
region is hottest just below the shock front
and then cools and compresses as it falls toward the surface \citep{2000NewAR..44...57C}.
The energy output from cyclotron radiation has a stronger temperature dependence
than that from bremsstrahlung,
so we expect to see the optical cyclotron emission from higher up
in the post-shock region where the gas is hotter.
Also the power radiated via bremsstrahlung depends quadratically on
the density while that for cyclotron radiation does so only linearly.
Therefore bremsstrahlung is preferably emitted from a lower height
where the gas is more compressed.
Because of the lower temperatures and higher densities at the bottom
of the post-shock region it is also expected that the soft X-rays
are produced at a lower height than the hard X-rays.

The above method to disentangle size and height is fairly accurate
if the thickness of the emission region is small compared to its height.
Even if this is not the case, equation (\ref{equ2}) still yields
a good estimate for the average height.
However, the size of a homogeneous emission region extending from height $H_1$ to $H_2$
will be overestimated by an angle $\sim$400$^\circ\cdot(\sqrt{H_2}-\sqrt{H_1})\cdot\sin^2i\cdot\sin^2\delta$.
For instance a thickness of $0.005\ R_{WD}$ and a height of $0.01\ R_{WD}$
will lead to a deviation of $\sim$10$^\circ$.
Another systematic error might be introduced into the height determination,
if the emission region becomes optically thick when viewed edge-on.
In this case the width of the bright phase is smaller
than expected and the height would be underestimated.
Further studies of the light curve profiles should certainly
include 3-dimensional modeling of the temperature and
density distributions in the post-shock region
and take into account optical depth effects
\citep{2000NewAR..44...57C,1983ApJ...268..291I}.

From observations done in 1987, \citet{1988MNRAS.234P..19B} found
a bright phase width for the cyclotron component of $\Delta\Phi=0.585$.
This is significantly larger than our measurement of $\Delta\Phi=0.551$.
We could attribute this discrepancy to a change in the height of the shock.
With the 1987 value for $\Delta\Phi$ we find a height of $\sim$0.019--0.040
which is about 3 times larger than during the \xmm\ observation.
This is surprising since in 1987 the total accretion rate was higher (see Section \ref{wwaccstate})
and according to current models the shock should have been lower.
It is possible that in 1987 the accretion spot was significantly
larger so that the accretion rate per area was actually lower
than during the \xmm\ observation,
which would lead to a larger height of the shock.
The difference between the two observations might also be due
to a change in the colatitude of the accretion region.
However, this is less likely since the orbital inclination near $90^\circ$
would require a very large colatitude change.


\section{Conclusions}

We have performed extensive \xmm\ observations of two
magnetic cataclysmic variable stars, DP Leo and WW Hor.
From a comparison with previous observations we conclude
that both objects were in an intermediate state of accretion.
We derived for both systems the mass ratio, orbital inclination
and an improved eclipse ephemeris.
We examined the X-ray data for periodic and quasi-periodic oscillations,
but found no evidence for them in either source.

\subsection{DP Leo}

From the shape of the hard X-ray light curves we conclude
that the post-shock region has a significant optical depth.
Although the X-ray spectrum indicates optically thin bremsstrahlung,
it might still be consistent with an optically thick post-shock
region if electron scattering is the major source of opacity.
We find a strong asymmetry in the shape of the soft X-ray light curve
which is not seen for the hard X-rays.
This asymmetry can be explained
with photoelectric absorption by the accretion curtain ($N_H\sim 10^{20}{\rm cm}^{-2}$)
or with two additional accretion regions near the lower rotational pole.
By comparing our measurement of the accretion spot longitude with previous results,
we find that the longitude is changing linearly in time by $2.3^\circ$ per year.
This is likely due to the rotational period of the white dwarf
being shorter than the orbital period by $\sim$1$\times10^{-6}\ $s$\ $s$^{-1}$.
From a comparison with previous results we find that our measurement
of the time of eclipse is not consistent with a linear ephemeris
and that instead the orbital period of the system is decreasing on a time scale
$P_{orb}/\dot{P}_{orb}=-3.3\cdot10^7\ $yrs.
This change is one order of magnitude faster than expected
for energy loss by gravitational radiation alone.

\subsection{WW Hor}

We use the shape of the X-ray light curves to derive
the horizontal and vertical extent of the post-shock region.
We find that the soft X-rays originate from a region that has a longitudinal extent of $\sim$40$^\circ$
while the hard X-rays are emitted in a region about half this size.
This result shows that the temperature of the post-shock gas is
increasing toward the centre.
We find that the cyclotron radiation originates from a height of $\sim$1\% of the white dwarf radius
while the X-rays are emitted at a much lower height.
This is consistent with the general prediction that
the temperature decreases and the density increases as the gas settles onto the surface.
By comparing our measurement with previous results we find that
the accretion longitude is correlated with the accretion rate.
This indicates that the observed longitude variations are due to a changing
location of the accretion spot
and not a change in the relative orientation of the magnetic axis.


\section*{acknowledgements}

This work is based on observations obtained with \xmm,
an ESA science mission with instruments and contributions
directly funded by ESA Member States and the USA (NASA).
D.P. acknowledges support from NASA grant NAG5-7714.
We thank Jason Etherton for the optical data of DP Leo
he obtained with the JKT (La Palma).


\bibliographystyle{mn2e}
\bibliography{polar}

\label{lastpage}

\end{document}